\documentclass[]{iopart}
\usepackage{graphics}
\usepackage{graphicx}
\usepackage{amsfonts}
\usepackage{amssymb}
\usepackage{cite}
\usepackage{xcolor}
\usepackage{dsfont}
\usepackage[latin1]{inputenc}
\usepackage{amsthm}
\usepackage{amsmath}
\usepackage{amscd}

\theoremstyle{plain}

\def\beqa{\begin{eqnarray}}
\def\eeqa{\end{eqnarray}}

\def\hw{\hbar \omega}
\def\hw4{ \frac {\hbar \omega}{4}}

\def\uni{{\bf i}}
\def\re{{\rm {e}}}

\def\b0{b_0}

\def\dag{^\dagger}
\def\pt{{\mathcal{PT}}}
\def\nonn{\nonumber \\}

\begin{document}

\title{Non-standard quantum algebras and finite dimensional $\mathcal{PT}$-symmetric systems}

\author{ \'Angel Ballesteros$^{1}$, Romina Ram\'\i rez$^{2}$ and Marta Reboiro$^{3}$ }

\address{{\small\it $^{1}$} Department of Physics, Universidad de Burgos, Burgos, Spain}
\address{{\small\it $^{2}$} IAM, CONICET-CeMaLP, University of La Plata, Argentina}
\address{{\small\it $^{3}$}IFLP, CONICET-Department of Physics, University of La Plata, Argentina}

\ead{reboiro@fisica.unlp.edu.ar}
\ead{romina@mate.unlp.edu.ar}
\ead{angelb@ubu.es}

\vspace{10pt}
\begin{indented}
\item[]September 26
\end{indented}

\begin{abstract}
In this work, $\mathcal{PT}$-symmetric Hamiltonians defined on quantum $sl(2, \mathbb R)$ algebras are presented. We study the spectrum of a family of non-Hermitian Hamiltonians written in terms of the generators of the non-standard $U_{z}(sl(2, \mathbb R))$ Hopf algebra deformation of $sl(2, \mathbb R)$. By making use of a particular boson representation 
of the generators of $U_{z}(sl(2, \mathbb R))$, both the co-product and the commutation relations of the quantum algebra are shown to be invariant under the $\mathcal{PT}$-transformation. In terms of these operators, we construct several finite dimensional $\mathcal{PT}$-symmetry Hamiltonians, whose spectrum is analytically obtained for any arbitrary dimension. In particular, 
we show the appearance of Exceptional Points in the space of model parameters and we discuss the behaviour of the spectrum both in the exact $\mathcal{PT}$-symmetry and 
the broken $\mathcal{PT}$-symmetry dynamical phases. As an application, we show that this non-standard quantum algebra can be used to define an effective model Hamiltonian describing accurately the experimental spectra of three-electron hybrid qubits based on asymmetric double quantum dots. Remarkably enough, in this effective model, the deformation parameter $z$ has to be identified with the detuning parameter of the system.
\end{abstract}

%
\vspace{2pc}

%

\section{Introduction}

Since the celebrated work of Bender and Boetcher \cite{bender0}, the study of mathematical properties \cite{bender1,bender2,bender3,bender4,bender5,bender6,ali} and physical applications \cite{bender8,pt1} of non-hermitian Hamiltonians has grown exponentially. Among non-hermitian operators,  those obeying Parity-Time Reversal symmetry ($\mathcal{PT}$-symmetry) have received particular attention, due to the rich structure that both its spectrum and dynamics \cite{ali} do present. A characteristic feature of these Hamiltonians is that its space of model parameters consists of two regions of well-defined structure: A region with real spectrum, where the eigenvectors of the Hamiltonian are also eigenvectors of the $\pt$-symmetry operator (the so-called exact $\pt$-symmetry phase) and another region where the spectrum includes complex pair conjugate eigenvalues (the broken $\pt$-symmetry phase), where the eigenstates of the Hamiltonian are not eigenstates of the $\pt$-symmetry operator. The boundary between these two regions is formed by the so-called Exceptional Points (EPs) \cite{eps1,eps2,epstheo}. At these latter values of the parameters of the model, two or more eigenvalues are degenerated and their eigenstates are coalescent. The consequence of the existence of EPs has been intensively analysed both theoretically \cite{eps1,pt1,epstheo} and experimentally \cite{eps2,pt1}.


In this work, we investigate the features of the spectra of a family of $\pt$-symmetry Hamiltonians defined on a non-standard Hopf algebra deformation of $sl(2, \mathbb R)$ \cite{uzfirst,BH1,BH2,BH3,BH4}, whose finite-dimensional boson representations were constructed in \cite{BH1,BH2}. By suitably modifying these representations we will be able to establish a $\pt$-invariant realisation of the $U_z(sl(2,\mathbb R))$ Hopf algebra. Therefore, in terms of these operators, we will be able to construct a family of $\mathcal{PT}$-symmetric Hamiltonians whose spectra will be analysed. Moreover, we shall show that this realisation of the quantum algebra $U_z(sl(2,\mathbb R))$ is physically sound since it can be used to define effective Hamiltonians that reproduce the structure of the eigenvalues that have been recently obtained for Hamiltonian models describing realistic systems of semiconductor quantum dots.

The paper is organised as follows. In Section \ref{form} we review the essentials of the non-standard $U_z(sl(2, {\mathbb R}))$ quantum algebra and find a boson representation for which its Hopf algebra structure is fully invariant under $\pt$-symmetry transformations. In Section \ref{results}, we construct two different families of Hamiltonians in terms of the $\pt$-symmetric generators of $U_z(sl(2,\mathbb R))$. Afterwards, we study their spectra by introducing similarity transformations for them to obtain isospectral Hamiltonians, and we discuss the regions in the space of parameters showing $\pt$-symmetry and broken $\pt$-symmetry, as well as the associated sets of EPs. As an application of these non-hermitian Hamiltonians, in Section {\ref{appli}} we introduce a suitable choice of the Hamiltonian and its parameters that reproduces the spectra of a system of three electrons in an asymmetric two-dimensional double well, which has been recently implemented experimentally \cite{ejQbit3}. Conclusions and outlook are given in Section \ref{con}. 

\section{Formalism} \label{form}

In this section, after reviewing the essentials of the non-standard $U_z(sl(2,\mathbb R))$ quantum Hopf algebra \cite{uzfirst,BH1,BH2,BH3,BH4}, we shall introduce a boson realisation of this algebra that is shown to be invariant under the $\pt$-transformation. This will be the realisation used in the rest of the paper in order to define new non-hermitian and $\pt$-symmetric Hamiltonian models.

\subsection{Dyson's boson representation of $sl(2,\mathbb R)$}

Let us begin by summarising the properties of the $sl(2, R)$ Lie algebra (see, for instance, \cite{gilmore}) whose generators, $\{L_{0}, L_{+}, L_{-}\}$, obey the well known commutation relations
\begin{equation}\label{comm0}
[L_{0},L_{\pm}]=\pm 2 L_{\pm} \ \ \ \ \ \, [L_{+},L_{-}]=  L_{0} \, .
\end{equation}
We point out that $\pt$-symmetric realisations of $sl(2,R)$ can be constructed in terms of boson operators, e.g.
\begin{equation}
L_{+} = -{\bf i}~ a^{\dagger} \ \ \ \ \ L_{0} = 2 ~a^{\dagger} a + \beta I \ \ \ \ \  L_{-}
 =-{\bf i}~ (a^{\dagger}a+\beta) ~a.
\label{sl2}
\end{equation}
where $a^{\dagger}$ is the creation operator, $a$ the annihilation operator and $\beta$ a free real parameter which is directly related with the eigenvalue of the Casimir operator $C = \frac{1}{2} L_{0}^{2}+ L_{+}L_{-} + L_{-}L_{+}$. 

The realisation \eqref{sl2} is a modification of the Gelfan'd-Dyson (GD) one-boson realisation \cite{GD1,GD2}. It can be straightforwardly shown that \eqref{sl2} is invariant under the $\mathcal{PT}$-transformation, since $ (\mathcal{PT})A(\mathcal{PT})^{-1}= -A $ for $A \in \{a, a^{\dagger}, {\bf {i}}\}$. 
As a consequence, many $sl(2, R)$ Hamiltonian systems constructed in terms of the bosonic mapping \eqref{sl2} of the generators of $sl(2, R)$ will obey $\mathcal{PT}$-symmetry. 

As an instructive example, we shall study the properties of the spectrum of the linear Hamiltonian 
\beqa
H_{\mu}= \mu L_{-}+ L_{+}.
\label{hnd}
\eeqa
It is straightforward to prove that for real $\mu$, this Hamiltonian is invariant under a $\mathcal{PT}$-transformation. Following \cite{fring1,fring2}, in order to find the associated spectrum, we can introduce a similarity transformation $\eta=\re^{\alpha L_{+}}e^{\beta L_{-}}e^{\gamma L_{z}}$ such that $h_\mu= \eta H_{\mu} \eta^{-1}$. In the following Proposition, we state the characteristics of the spectrum of $h$ depending on the sign of $\mu$.\\

\noindent 
{\bf{Proposition 1.}}
Let $\eta=e^{\alpha L_{+}}e^{\beta L_{-}}e^{\gamma L_{z}}$. 

\noindent a) If
$\alpha_{0} =-\frac{e^{2 \gamma}}{2 \sqrt{\mu }} $ and $\beta = e^{-2 \gamma_{0} } \sqrt{\mu }$, then $h_{\mu}=\eta H_{\mu} \eta^{-1}$ is hermitian for $\mu >0$.

\noindent b)
For $\mu<0$, $h_{\mu}$ is a diagonal matrix with complex pair-conjugate eigenvalues.

\begin{proof} By using the relations presented in \ref{ap2}, we have
\begin{eqnarray*}
h_\mu & = &\eta(\mu L_{-}+L_{+})\eta^{-1} \nonn
&= & L_{-}e^{-2 \gamma } \left(\mu -\beta ^2 e^{4 \gamma
   }\right)+
L_{+} \left(e^{2 \gamma } (\alpha  \beta
   +1)^2-\alpha ^2 e^{-2 \gamma } \mu \right)+\nonn
&&+L_{z}
   \left(\alpha  e^{-2 \gamma } \mu -\beta  e^{2 \gamma }
   (\alpha  \beta +1)\right).
   \end{eqnarray*}

In the bosonic realisation \eqref{sl2},  $L_{z}$ is the only hermitian generator. Then, to find the isospectral hermitian operator associated with $H_{\mu}$ we need to cancel the terms including $L_{+}$ and $L_{-}$. Because of that, we obtain the equations
\begin{eqnarray*}
0&=& \left(e^{2 \gamma } (\alpha  \beta
   +1)^2-\alpha ^2 e^{-2 \gamma } \mu \right) \, ,\\
0&=& e^{-2 \gamma } \left(\mu -\beta ^2 e^{4 \gamma
   }\right) \, ,
\end{eqnarray*}
whose solution is $(\alpha_{0}, \beta_{0}) =(-\frac{e^{2 \gamma_{0} }}{2 \sqrt{\mu }}, e^{-2 \gamma_{0} } \sqrt{\mu })$.
In that case,
\beqa
h_\mu= \eta(\mu L_{-}+L_{+})\eta^{-1}=\sqrt{\mu}L_{0}=\sqrt{\mu}(2 a^{\dagger} a + \beta I).
\eeqa
Thus, $h_\mu$ has real spectrum for $\mu>0$ and has complex pair-conjugate eigenvalues for $\mu <0$.
\end{proof}

\subsection{The non-standard quantum algebra $U_z(sl(2, {\mathbb R}))$}

Many authors have considered different deformations of the algebra $sl(2, R)$ and have applied them in different contexts (see, for instance, \cite{higgs,debergh1,debergh2,polynos}).
We recall that among all possible deformations of a given Lie algebra, a distinguished class is defined by the Hopf algebra deformations of the Universal Enveloping Algebra (UEA) of such Lie algebra. These deformed Hopf algebras are called Quantum Universal Enveloping Algebras (QUEA) or, in short, quantum algebras, and are defined as simultaneous and compatible deformations of both the commutation rules of the Lie algebra and the coproduct map that defines its tensor product representations \cite{ChariPressley,Majid}.

In the case of $sl(2, R)$, we will deal with the so-called non-standard quantum deformation $U_z(sl(2, {\mathbb R}))$ \cite{BH1,BH2} (the standard deformation is the Drinfel'd-Jimbo one \cite{Dr,Ji}).  
Its generators, named $\{j_{0}^{(z)},j_{+}^{(z)}, j_{+}^{(z)}\}$, where $z$ is a real deformation parameter, define the quantum algebra relations through the commutation rules
\begin{equation}\label{commz}
[j_{0}^{(z)},j_{+}^{(z)}]= \tfrac{e^{2 z j_{+}^{(z)}}-1}{z} \ \ \ \ [j_{0}^{(z)},j_{-}^{(z)}] = -2 j_{-}^{(z)}+z (j_{0}^{(z)})^{2} \ \ \ \ [j_{+}^{(z)},j_{-}^{(z)}]= j_{0}^{(z)}\, ,
\end{equation}
which are just a generalisation of \eqref{comm0}, which is smoothly recovered in the $z\to 0$ limit.

Tensor product representations of the quantum algebra \eqref{commz} are obtained through the so-called coproduct map
\begin{eqnarray}
\Delta ({    {j}}_0^{(z)}) & = & 1 \otimes {    {j}}_0^{(z)} + {    {j}}_0^{(z)} \otimes \re^{2 z {    {j}}_+^{(z)} }, \nonumber \\
\Delta ({    {j}}_-^{(z)}) & = & 1 \otimes {    {j}}_-^{(z)} + {    {j}}_-^{(z)} \otimes \re^{2 z {    {j}}_+^{(z)} }, \nonumber \\
\Delta ({    {j}}_+^{(z)}) & = & 1 \otimes {    {j}}_+^{(z)} + {    {j}}_+^{(z)} \otimes 1, 
\label{hopf0}
\end{eqnarray}
which defines an algebra homomorphism between $U_z(sl(2, {\mathbb R}))$ and $U_z(sl(2, {\mathbb R}))\otimes U_z(sl(2, {\mathbb R}))$. As expected, the limit $z\to 0$ leads to the usual (undeformed) rule for the construction of $sl(2, R)$ tensor product representations.

We are interested in getting the non-standard deformed generalisation of the ${\mathcal{PT}}$-symmetric GD realisation \eqref{sl2}. Such a result can be obtained by starting from the $U_z(sl(2, {\mathbb R}))$ boson representation obtained in \cite{BH2,BH1}, together with the definition of the new set of operators,
$\{     {J}_{0}^{(z)},    {J}_{+}^{(z)},     {J}_{-}^{(z)} \}$ as
 \beqa
    {J}_{0}^{(z)}&=&~~j_{0}^{(-\uni z)}, \nonumber \\
    {J}_{\pm}^{(z)} &=& \mp \uni~ j_{\pm}^{(-\uni z)}\, .
\label{genpt}
\eeqa
In such a way we obtain
\begin{eqnarray}
    {J}_{+}^{(z)} & = & = -\uni a^{\dag} , \nonumber \\  
    {J}_{0}^{(z)} & = & 
\uni \tfrac{\re^{-2 \uni z a^{\dag}} - 1}{z} a + \beta \tfrac{ \re^{-2 \uni z a^{\dag}} + 1}{2}, \label{defboson} \\
    {J}_{-}^{(z)} &=& \tfrac{\re^{-2 \uni z a^{\dag}} -1}{2  z}a^{2}- \uni \beta \tfrac{\re^{-2 \uni z a^{\dag}} + 1}{2} a -z \beta^{2}\tfrac{\re^{-2 \uni z a^{\dag}} -1}{8}\, .
\nonumber   
\end{eqnarray}
These operators can be straightforwardly shown to be ${\mathcal{PT}}$-symmetric, and we stress that the transformation $z\rightarrow -\uni z$ is essential in order to recover the ${\mathcal{PT}}$ symmetry of this boson representation of the deformed algebra.

The action of the operators $\{{J}_{0}^{(z)},{J}_{+}^{(z)},{J}_{-}^{(z)} \}$ on the eigenstates $\{ |m \rangle , (m=0,1,\dots\infty)\}$ of the usual boson number operator $a^{\dag}a$, provides their lower-bounded representation, namely

\begin{flalign}
 &    {J}_{+}^{(z)}| m\rangle = -\uni \sqrt{m+1}|m+1\rangle, \nonumber \\    
 &    {J}_{0}^{(z)}| m \rangle = 
    (2 m + \beta) | m \rangle \nonn
 & ~~~~~~~~~~~~+ 
   \sum_{k \geq 1} \frac{(- 2 \uni z)^{k}}{k!}\sqrt{\frac{(m+k)!}{m!}} \left(\frac{2m}{k+1}+\frac{\beta}{2} \right) |m+k\rangle , \nonn
 &   {J}_{-}^{(z)}| m\rangle= - \uni \sqrt{m}(m-1+\beta) |m-1\rangle
\nonn 
 & ~~~~~~~~~~~~-\uni \sum_{k \geq 1} \frac{(- 2 \uni z)^{k}}{k!}\sqrt{\frac{(m+k)!}{m!}} 
\left [ \frac{m}{\sqrt{m+k}} \left(\frac{m-1}{k+1}+\frac{\beta}{2} \right) |m-1+k\rangle 
\right. \nonumber\\
& ~~~~~~~~~~~~~~~~~~~~~~~~~~~~~~~~~~~~~~~~~~~~~~~~~~~~~~~~~~~~~~~~~~~~~~~~~~~~~~
\left.- \uni z \frac{\beta^{2}}{8} |m+k\rangle \right ]. 
\label{melements} 
\end{flalign}

As it was shown in \cite{BH2}, for values of the parameter $\beta \in {\mathbb Z}^-$, this representation becomes reducible and leads to the ${\mathcal{PT}}$-symmetric finite-dimensional irreducible representations of dimension $d=|\beta-1|$ of the quantum algebra $U_z(sl(2, {\mathbb R}))$.

 The commutation rules, coproduct ($\Delta$),  counit ($\epsilon$), and antipode ($\gamma$) maps defining the full Hopf algebra structure of $U_z(sl(2, {\mathbb R}))$ in terms of the ${\mathcal{PT}}$-symmetric generators indeed have the same structure as those of $\{{j}_{0}^{(z)},{j}_{+}^{(z)},{j}_{-}^{(z)} \}$ (see \ref{ap1}). Namely,
\begin{eqnarray}
\Delta ({    {J}}_0^{(z)}) & = & 1 \otimes {    {J}}_0^{(z)} + {    {J}}_0^{(z)} \otimes \re^{2 z {    {J}}_+^{(z)} }, \nonumber \\
\Delta ({    {J}}_-^{(z)}) & = & 1 \otimes {    {J}}_-^{(z)} + {    {J}}_-^{(z)} \otimes \re^{2 z {    {J}}_+^{(z)} },  \\
\Delta ({    {J}}_+^{(z)}) & = & 1 \otimes {    {J}}_+^{(z)} + {    {J}}_+^{(z)} \otimes 1, \nonumber \\
\epsilon(X)& = & 0, ~~~X \in \{     {J}_{0}^{(z)},    {J}_{+}^{(z)},     {J}_{-}^{(z)} \},  \\
\gamma (    {J}_{0}^{(z)}) &=& -{    {J}}_{0}^{(z)} \re^{-2 z {    {J}}_{+}^{(z)}}, \nonumber \\
\gamma (    {J}_{-}^{(z)}) &=& -{    {J}}_{-}^{(z)} \re^{-2 z     {J}_{+}^{(z)} },  \\
\gamma (    {J}_{+}^{(z)}) &=& -{    {J}}_{+}^{(z)}, \nonumber
\label{hopf}
\end{eqnarray}
\beqa
~ [     {J}_{0}^{(z)},    {J}_{+}^{(z)}] & = & \frac{\re^{2 z     {J}_{+}^{(z)}}-1}{z}, \nonumber \\
~ [    {J}_{0}^{(z)},    {J}_{-}^{(z)}] & = & -2     {J}_{-}^{(z)}+z \left(     {J}_{0}^{(z)} \right)^2,  \label{commzpt}\\
~ [    {J}_{+}^{(z)},    {J}_{-}^{(z)}] & = &     {J}_{0}^{(z)}. \nonumber
\eeqa
Moreover, the deformed Casimir operator is given by:
\begin{flalign}
& C_z= \frac 12 J_0^{z} \re^{-2 z}J_+^{(z)} + \frac{1-\re^{-2 z {J}_{+}^{(z)}}}{2 z} J_- + J_- \frac{1-\re^{-2 z {J}_{+}^{(z)}}}{2z}+ \re^{-2 z {J}_{+}^{(z)}}-1,
\label{cas}
\end{flalign}
and its eigenvalues are expressed in terms of $\beta$ as $C=\beta(\beta/2-1)$.

Finally, it can be easily proved (see \ref{ap1}) that both the commutation relations given in \eqref{commz} and the coproduct map are preserved under $\pt$ symmetry transformations, which means that
\begin{eqnarray}
(\pt) \Delta ({    {J}}_0^{(z)}) (\pt)^{-1}& = & \Delta ({    {J}}_0^{(z)}), \nonumber \\
(\pt) \Delta ({    {J}}_+^{(z)}) (\pt)^{-1}& = & \Delta ({    {J}}_+^{(z)}), \nonumber \\
(\pt) \Delta ({    {J}}_-^{(z)}) (\pt)^{-1}& = & \Delta ({    {J}}_-^{(z)}).
\end{eqnarray}

In the rest of the paper, we shall apply the previous results to the study of a $\pt$-symmetric family of Hamiltonians obtained from the finite-dimensional irreducible representation of the $\pt$-invariant generators \eqref{defboson} of the non-standard $U_z(sl(2,\mathbb R))$ Hopf algebra.

\section{Results and discussion} \label{results}

In this Section, we present a large family of $\pt$-symmetric Hamiltonians defined as functions of the operators (\ref{genpt}) under the realisation \eqref{defboson}. We will show the appearance of Exceptional Points in the space of model parameters and we will discuss the behaviour of the spectrum both in the exact $\mathcal{PT}$-symmetric and the broken $\mathcal{PT}$-symmetric dynamical phases. 

As an initial step in the understanding of the problem, we shall study the natural deformed generalisation of the Hamiltonian of \eqref{hnd}, namely the operator
\beqa
H_{\mu}=\mu J_{-}^{(z)}+J_{+}^{(z)}.
\label{h1}
\eeqa
If we consider the representation of dimension 2 of the generators (this means \eqref{melements} with $\beta=-1$), given by
\begin{flalign}
& J_{+}^{(z)}= 
\left(
\begin{array}{cc}
 0 & 0  \\
 - \uni & 0 \\
\end{array}
\right), \ \ \ J_{-}^{(z)}= 
\left(
\begin{array}{cc}
 0 & \uni  \\
\frac{\uni z^2}{4} & z \\
\end{array}
\right),  \ \ \ J_{0}^{(z)}= 
\left(
\begin{array}{cc}
 -1 & 0  \\
 \uni z & 1 \\
\end{array}
\right),
\label{2drep}
\end{flalign}
the Hamiltonian of \eqref{h1} reads
\beqa
H_\mu= 
\left(
\begin{array}{cc}
 0 & \uni \mu  \\
 \frac{1}{4} \uni z^2 \mu - \uni & z \mu  \\
\end{array}
\right).
\eeqa
Indeed, in this case, analytical results can be obtained: As the Hamiltonian of \eqref{h1} obeys $\pt$-symmetry,  we can construct an operator $S$ such as $S H= H^{\dagger}S$. For instance
\beqa
S=\left(
\begin{array}{cc}
 \frac{z^2}{2}+\frac{2}{\mu } & -\uni z \\
 \uni z & 2 \\
\end{array}
\right),
\eeqa
where the operator $S$ is self-adjoint and for $\mu>0$ is positive definite. It is now possible to construct a self-adjoint Hamiltonian through the similarity transformation $h_\mu=S^{1/2} \ H_\mu \ S^{-1/2}$, where
\beqa
h_\mu= \left(
\begin{array}{cc}
 \tfrac{z \mu  \left(\left(z^2+4\right) \mu -4\right)}{2 \left(\left(z^2+4\right) \mu +8 \sqrt{\mu }+4\right)} & -\tfrac{\uni \sqrt{\mu } \left(\left(z^2-4\right) \mu -8 \sqrt{\mu
   }-4\right)}{\left(z^2+4\right) \mu +8 \sqrt{\mu }+4} \\
 \tfrac{\uni \sqrt{\mu } \left(\left(z^2-4\right) \mu -8 \sqrt{\mu }-4\right)}{\left(z^2+4\right) \mu +8 \sqrt{\mu }+4} & \tfrac{z \mu  \left(\left(z^2+4\right) \mu +16 \sqrt{\mu
   }+12\right)}{2 \left(\left(z^2+4\right) \mu +8 \sqrt{\mu }+4\right)} \\
\end{array}
\right).
\eeqa
For $\mu<0$, the operator $h_\mu$ is isospectral with respect to $H_\mu$ but it is not hermitian. Due to the fact that $H_\mu$ and $H_\mu^\dagger$ are similar operators, their spectrum is either real or contains complex pair conjugate eigenvalues. In this example, the eigenvalues take the form
\beqa
E_\pm = \frac {\mu} 2 z \pm \sqrt{\mu}. 
\eeqa

We also recall that in the treatment of Hamiltonians with $\mathcal{PT }$-symmetry, a similar transformation between $H$ and its adjoint can be obtained by constructing a bi-orthogonal basis \cite{RRK}. In fact, following \cite{RRK}, let $\mathcal{A}_{H}=\{\widetilde{\phi}_{j}\}_{j=1...N_{max}}$ the eigenvectors of a non-hermitian operator $H$, i.e $H\widetilde{\phi}_{j} = \widetilde{E}_{j}\widetilde{\phi}_{j}$. In the same way we denote $\mathcal{A}_{H^{\dagger}}=\{\overline{\psi}_{j}\}_{j=1...N_{max}}$ as the eigenvectors of $H^{\dagger}$, and therefore $H^{\dagger}\overline{\psi}_{j} = \overline{E}_{j}\overline{\psi}_{j}$. If H is diagonalisable, the sets  $\mathcal{A}_{H}$ and $\mathcal{A}_{H^{\dagger}}$ form a bi-orthonormal set of $H$,
i.e. $ \langle \overline{\psi}_{i} | \widetilde{\phi}_{j} \rangle = \delta_{ij}$ and $\widetilde{E}_{j}=\overline{E}_{j}^{*} $.
Following \cite{RRK}, for a pseudo-hermitian diagonalisable Hamiltonian with real spectrum we can define a symmetry
operator $S_{\psi}$ so that $S_{\psi}H=H^{\dagger}S_{\psi}$. Moreover, in terms of $\overline{\psi}_{j}$, the operator $S_{\psi}$ is given by
\beqa 
S_{\psi}=\sum_{j=1}^{N_{max}}| \overline{\psi}_{j}\rangle \langle \overline{\psi}_{j} |.
\label{sym}
\eeqa
When this operator is self-adjoint and positive definite,  an inner product can be implemented as  $\langle f | g \rangle_{S_{\psi}} =\langle f | S_{\psi}g \rangle$. If $S$ is not positive definite we are then forced to work within the framework of the formalism of Krein spaces, as it has been described in detail in \cite{RRK}. 


It is also worth mentioning that, in general, the constant $\mu$ generates only a scalar distortion in the spectrum of the Hamiltonian \eqref{h1}. Therefore, according to the sign of the coupling constant $\mu$, we can make a change of variables that allow us to restrict our study to the two Hamiltonians $H_{1}$ and $H_{-1}$.
\\

\noindent 
{\bf{Proposition 2.}}
Hamiltonians $H_{\mu}=\mu J_{-}^{(z)}+J_{+}^{(z)}$ can be mapped into $H_{1}$ and $H_{-1}$ for $\mu>0$ and $\mu<0$, respectively. 

\begin{proof}
If $\mu>0$
\begin{eqnarray*}
H_{\mu}&=& \mu J_{-}^{(z)}+J_{+}^{(z)}\\
&=& \sqrt{\mu} \left(\sqrt{\mu} J_{-}^{(z)}+\frac{1}{\sqrt{\mu}} J_{+}^{(z)} \right) \, .
\end{eqnarray*}
Through the change of parameters $$\lambda:= \sqrt{\mu} z \ \ \ \ \ b_{-}=\sqrt{\mu}{a}\ \ \ \ \ b_{+}=\frac{1}{\sqrt{\mu}}a^{\dagger}$$ we find a new bosonic representation given by $\{b_{+},b_{-} \}$ and a rescaled deformation parameter called $\lambda$ such that $H_{\mu}$ is rewritten as 
$$H_{1}=J_{-}^{(\lambda)}+J_{+}^{(\lambda)} \, .$$

In the same way, if $\mu<0$, we take $-\mu=  \nu >0$ and 
\begin{eqnarray*}
H_{-\nu}&=& -\nu J_{-}^{(z)}+J_{+}^{(z)}\\
&=& -\sqrt{\nu} \left(\sqrt{\nu} J_{-}^{(z)}-\frac{1}{\sqrt{\mu}} J_{+}^{(z)} \right) \, .
\end{eqnarray*}
With the new parameters 
\begin{equation}\label{lambda}
\lambda:= \sqrt{\nu} z \ \ \ \ \ b_{-}=\sqrt{\nu}{a}\ \ \ \ \ b_{+}=\frac{1}{\sqrt{\nu}}a^{\dagger}
\end{equation}
again the equivalence between $H_{\mu}$  and   $-H_{-1}=J_{-}^{(\lambda)}-J_{+}^{(\lambda)} $ can be established.
\end{proof}

As it could be expected, difficulties in finding an analytical expression for $h_\mu$ increase when we consider higher dimensional representations. In fact, the exact form of the spectrum of the Hamiltonian \eqref{h1} for an arbitrary finite-dimensional irreducible representation is not known. Nevertheless, the aim of this paper consists in presenting other families of Hamiltonians defined on $U_z(sl(2,\mathbb R))$ that can be exactly solved.\\

In particular, let us consider
the family of Hamiltonians given by
\beqa
H(\mu_+,\mu_-,\mu_0)=\mu_-~J_{-}^{(z)}+ \mu_+ [J_0^{(z)},{J}_{+}^{(z)}] + \mu_0 J_{0}^{(z)},
\label{hz}
\eeqa
with parameters $(\mu_+,\mu_-,\mu_0) \in \mathbb R$ and the commutator $[J_0^{(z)},{J}_{+}^{(z)}]$ given in \eqref{commz}.
Following \cite{fring1}, in order to characterise the spectrum we shall work with Hamiltonians which are isospectral partners of $H$. Therefore, let us introduce the similarity transformations $\Upsilon_\pm$ given by
\begin{flalign}
& \Upsilon_\pm = \re^{\eta~{J}_{0}^{(z)}}\re^{\kappa_\pm~{J}_{+}^{(z)}}, ~~~\kappa_\pm= \pm \frac{1}{\mu_-} \left ( \sqrt{\mu_0^2+ 2 \mu_+ \mu_-} -\mu_0 \right).
\end{flalign}

The transformed Hamiltonians can be obtained by using the expressions provided in \ref{ap2}, namely 
\begin{flalign}
& {\cal{H}}_\pm = \Upsilon_\pm \, H \,\Upsilon_\pm^{-1} 
=\mu_- \re^{-2 \eta} J_{-}^{(z)} + z \mu_- \re^{-\eta} \sinh(\eta) ~(J_0^{(z)})^2 \pm \sqrt{\mu_0^2+ 2 \mu_+ \mu_-}~J_0^{(z)}. 
\label{eqhcal}
\end{flalign}
For any value of the parameters $(\mu_+,\mu_-,\mu_0)$, the Hamiltonian ${\cal H}$ of \eqref{eqhcal} in the limit $\eta \rightarrow \infty$ reads 
\beqa
{\mathfrak h}_\pm(\mu_+,\mu_-,\mu_0) =  z~ \mu_- ~(J_0^{(z)})^2 \pm \sqrt{\mu_0^2+ 2 \mu_+ \mu_-}~J_0^{(z)} \, ,
\label{h}
\eeqa
and any finite dimensional irreducible representation of ${\mathfrak h}_\pm(\mu_+,\mu_-,\mu_0)$ of \eqref{h}, is given by a triangular matrix of dimension $d=|\beta-1|$. Then the following proposition can be proven through a straightforward computation:\\

\noindent {\bf{Proposition 3.}}
The Hamiltonians  ${\mathfrak h}_-(\mu_+,\mu_-,\mu_0) $,  and ${\mathfrak h}_+(\mu_+,\mu_-,\mu_0) $ of \eqref{h} are isospectral operators, and their spectrum $\sigma$ is given by 
\begin{equation}
\sigma =\left\{
              \begin{array}{ll}
              \tfrac{z}{2} ~\mu_{-} ~ (2k+1)^2  \pm  (2k+1) \sqrt{\mu_{0}^2+ 2 \mu_{+}\mu_{-}} , & \hbox{even  }d, ~~~ 0 \leq k \leq \frac{d-2}{2}\\
              \tfrac{z}{2} ~\mu_{-} ~ (2k)^2    \pm 2k \sqrt{\mu_{0}^2+ 2 \mu_{+}\mu_{-}},       & \hbox{~odd }d, ~~~ 0 \leq k \leq \frac{d-1}{2}.
              \end{array}
            \right.
\label{spectrum}
\end{equation}

As can be observed from \eqref{spectrum},
when $\mu_{0}+ 2 \mu_{+}\mu_{-}>0$, the spectrum of $h_\pm$ is real ({\em i.e.}, we are in the exact $\pt$-symmetry phase).
On the other hand, if $\mu_{0}+ 2 \mu_{+}\mu_{-}<0$, the spectrum consists of complex pair conjugate eigenvalues, thus indicating a broken $\pt$-symmetry phase. In the model space of parameters, the points at which $\mu_{0}+ 2 \mu_{+}\mu_{-}=0$ are EPs. These points provide the boundary between the two dynamical phases of the system.\\

It is worth noting that the Hamiltonian in \eqref{hz} may initially seem much more intricate compared to the Hamiltonian given in \eqref{h1}, given that the commutator involves the exponential of the operator $ J_{+}^{(z)} $. Nevertheless, it turns out that for this particular group of operators, it is possible to determine the spectrum explicitly by using similarity transformations, in contradistinction to what happens with the Hamiltonian $ \mu J_{-}^{(z)} + J_{+}^{(z)} $.  Moreover, the very same spectral analysis can be straightforwardly generalised to the family of operators
\begin{equation}
H_{g}(\mu_+,\mu_-,\mu_0)=\mu_-~J_{-}^{(z)}+ \mu_+ [J_0^{(z)},{J}_{+}^{(z)}] + \mu_0 g(J_{0}^{(z)}) \, ,
\end{equation}
with $g$ being a generic function of $J_{0}^{(z)}$.\\


In the following, we analyse the phase structure of the spectrum of the Hamiltonian ${\mathfrak h}_+(\mu_+,\mu_-,\mu_0)$ (\ref{h}).
As a first step, in order to discuss the appearance of EPs in the present model, 
let us take 
$|\mu_-|=-|\mu_+|=\mu$ and $\nu = \mu_0 /\mu$. In this manner we get 
\beqa
h_-(\mu,\nu) & = & {\mathfrak h}_+(-\mu_,\mu,\mu ~\nu)=\mu(z~(J_0^{(z)})^2 + \sqrt{\nu^2-2}~J_0^{(z)}).
\label{hm}
\eeqa

\begin{figure}
\includegraphics[width=12cm]{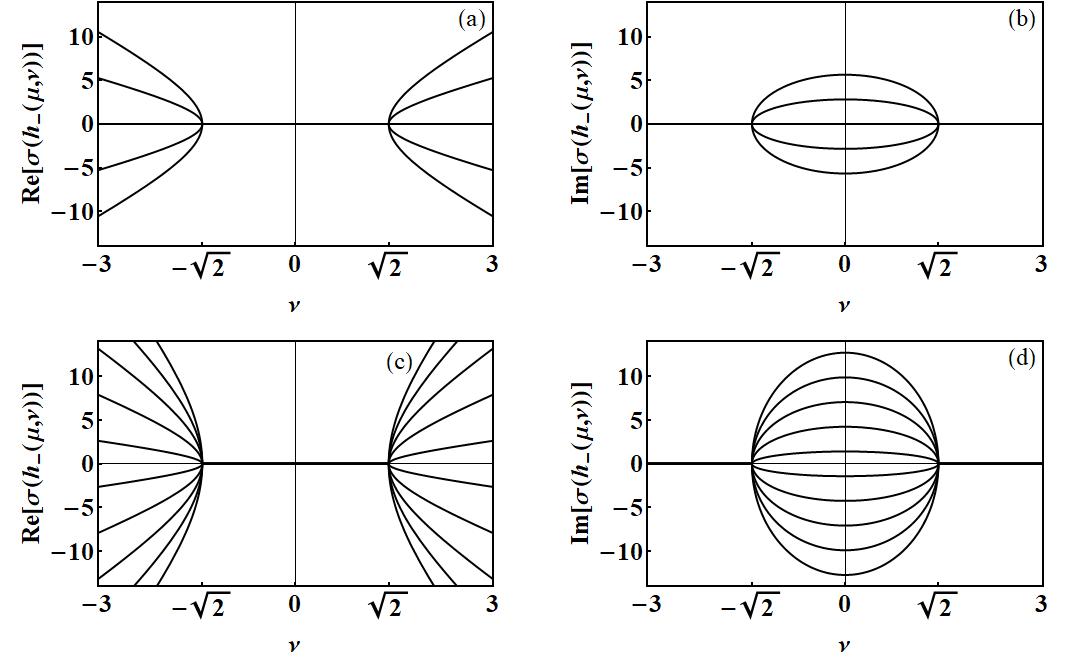}
\caption {Behaviour of the spectrum of $H_-(\mu, \nu)$ for $z=0$, as a function of $\nu$. The values are given in units of $\mu$. Panels (a) and (b) correspond to values of $\beta=-4$ (dimension $d=5$),  and Panels (c) and (d) correspond to values of $\beta=-9$ (dimension $10$). In Panels (a) and (c) we display the behaviour of the real part of the eigenvalues. In Panels (b) and (d), the imaginary part of the eigenvalues is depicted.}
\label{fig1}
\end{figure}

In Figure \ref{fig1}, we show the behaviour of its spectrum as a function of $\nu$, for the undeformed Hamiltonian with $z=0$. Panels (a) and (b) correspond to values of $\beta=-4$ (dimension $d=5$),  and Panels (c) and (d) correspond to values of $\beta=-9$ (dimension $10$). In Panels (a) and (c) we display the behaviour of the real part of the eigenvalues. In Panels (b) and (d), the imaginary part of the eigenvalues is depicted. It can be seen the presence of EPs of order 2 and 5 at $|\nu|=\sqrt{2}$, for the $d=5$ and $d=10$ cases, respectively.

\begin{figure}
\includegraphics[width=12cm]{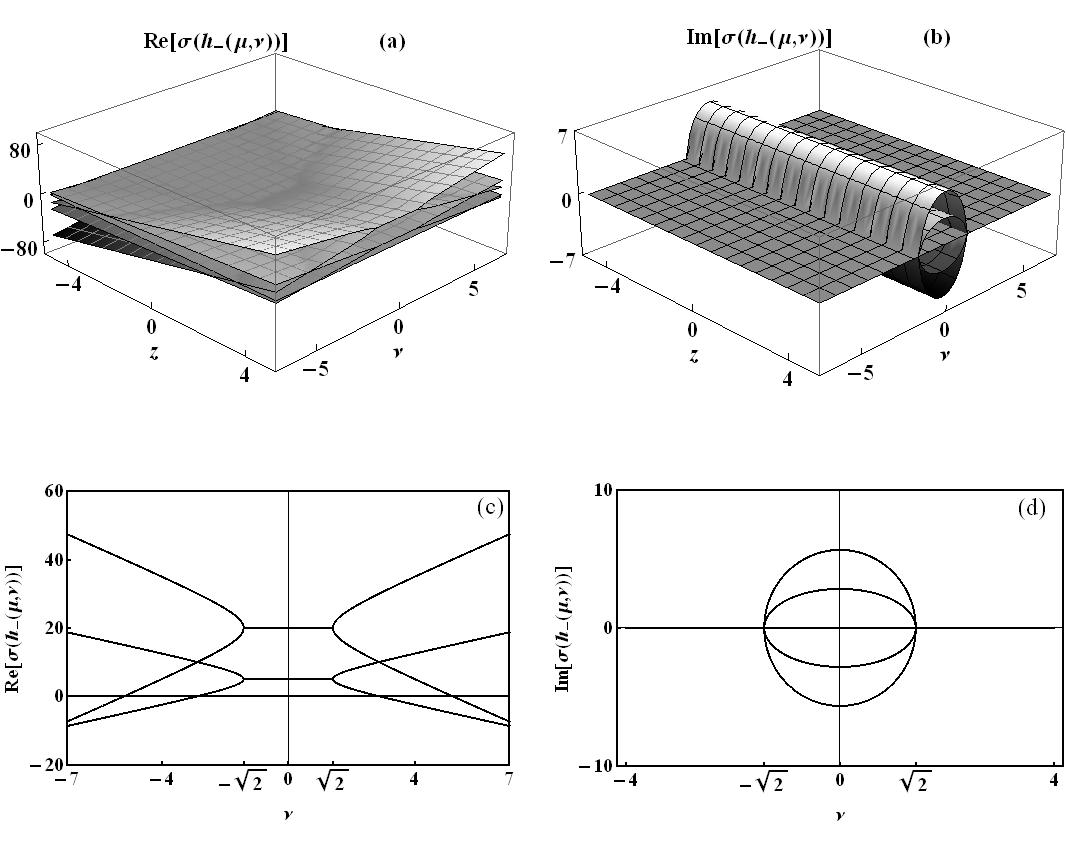}
\caption{Figure 2 shows the spectrum of the Hamiltonian $h_-(\mu,\nu)$ in units of $\mu$, as a function of $\nu$ and $z$, for dimensions $d=5$. In Panels (a) and (b) we plot the real and the imaginary part of the eigenvalues, respectively. 
In Panels (c) and (d), we present the projection for $z=2.5$. The real and imaginary parts of the eigenvalues are presented in (c) and (d), respectively.}
\label{fig2}
\end{figure}

\begin{figure}
\includegraphics[width=12cm]{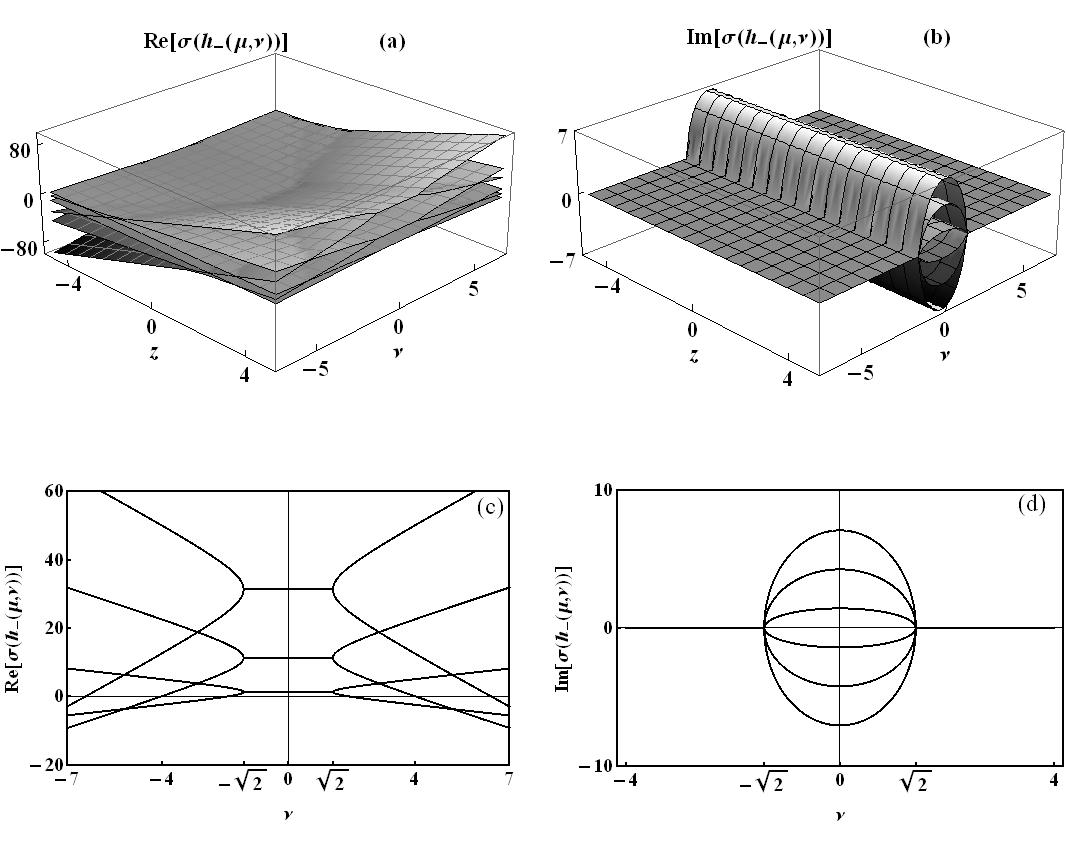}
\caption{The spectrum of the Hamiltonian $h_+(\mu,\nu)$ in units of $\mu$, as a function of $\nu$ and $z$, for dimensions $d=6$, is depicted in Figure 3. In Panels (a) and (b) we plot the real and the imaginary part of the eigenvalues, respectively. 
In Panels (c) and (d), we present a cut in the graph for $z=2.5$. The real and imaginary parts of the eigenvalues are presented in (c) and (d), respectively.}
\label{fig3}
\end{figure}

In Figures \ref{fig2} and \ref{fig3}, we display the spectrum of the deformed Hamiltonian $h_-(\mu,\nu)$ in units of $\mu$, as a function of both $\nu$ and the deformation parameter 
$z$, for dimensions $d=5$ and $d=6$. In Panels (a) and (b) we plot the real and the imaginary part of the eigenvalues, respectively. 
In Panels (c) and (d), we present the projection for the case $z=2.5$. The real and imaginary parts of the eigenvalues are presented in (c) and (d), respectively. The EPs occur at values of $\nu=\pm \sqrt{2}$.
For the Hamiltonian of \eqref{hm}, $h_-(\mu,\nu)$, the EPs lie between the region with exact $\pt$ symmetry (real spectrum) and the 
region of broken $\pt$-symmetry, with pairs of complex conjugate energies.\\

As a second example, let us consider  
$|\mu_-|=|\mu_+|=\mu$ and $\nu = \mu_0 /\mu$:
\beqa
h_+(\mu,\nu) & = &  {\mathfrak h}_+(\mu_,\mu,\mu ~\nu)= \mu(z~(J_0^{(z)})^2 + \sqrt{\nu^2+2}~J_0^{(z)}).
\label{hp}
\eeqa
In this case, the spectrum of $h_+(\mu,\nu)$, $\sigma(h_+(\mu,\nu))$, takes real values. 

\begin{figure}
\includegraphics[width=12cm]{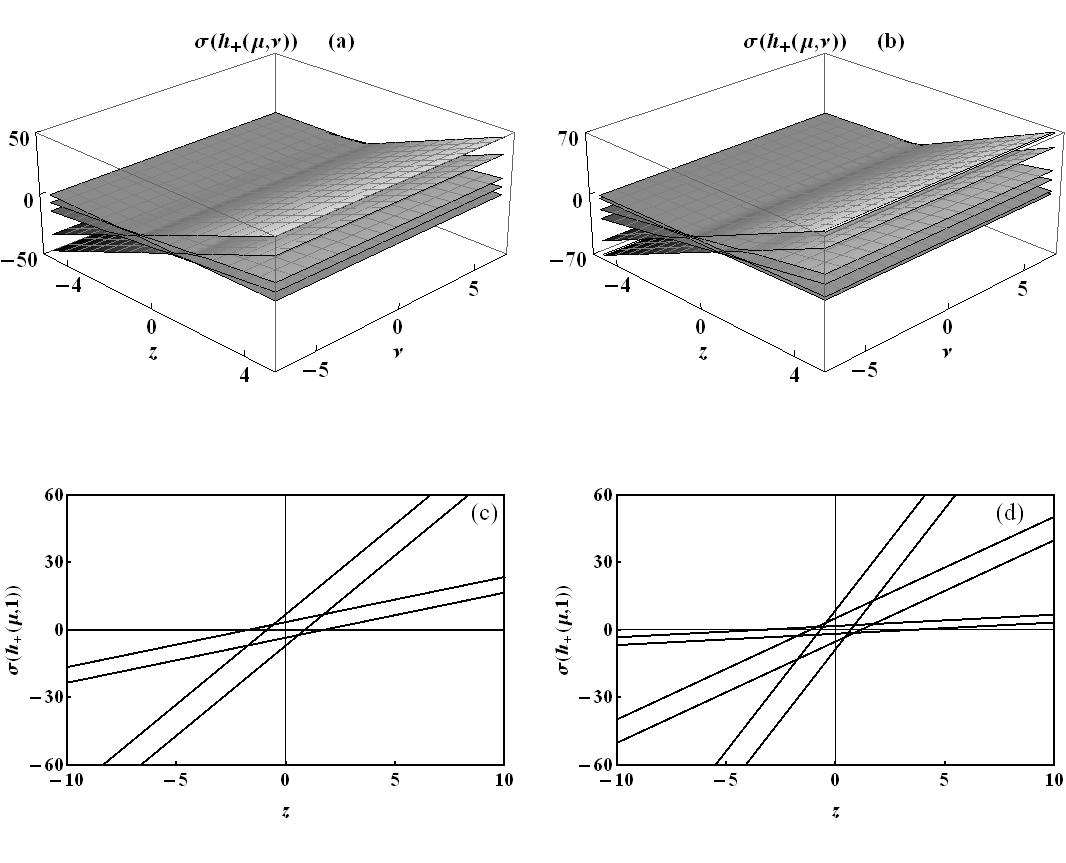}
\caption{We plot the spectrum of $h_+(\mu,\nu)$ in units of $\mu$, as a function of $\nu$ and $z$. Panels (a) and (c) correspond to the graphs for $d=5$ and Panels (b) and (d) to $d=6$. In Panels (c) and (d) we depict a projection of the plots (a) and (d) for $\nu=1$.  }
\label{fig4}
\end{figure}

In Figure \ref{fig4}, we plot the spectrum of $h_+(\mu,\nu)$ in units of $\mu$, as a function of $\nu$ and $z$. Panels (a) and (c) correspond to the results obtained for dimension $d=5$, while Panels (b) and (d) correspond to the results for dimension $d=6$. In Panels (c) and (d) we present the projections of the graphs at $\nu=1$, as a function of z. 
It is important to emphasise that, as a function of $z$, eigenvalues form `bands' composed of two energies. The distance between consecutive bands is governed by the second term of \eqref{spectrum}. This will be of the outmost relevance when dealing with the applications presented in the next Section.\\

Finally, we shall consider another family of exactly solvable Hamiltonians written in terms of the generators of $U_{z}(sl(2,\mathbb R))$, namely
\beqa
H_{0}= \mu_{-}J_{-}^{(z)}+ \sum_{n=1}^{N} a_{n} ~\left[ J_{0}^{(z)}\right]^{n}
\label{hz1}
\eeqa
where $N \in {\mathbb Z}^+ \cup \infty$.

As in the previous cases, by using the similarity transformation given now by the operator $e^{\alpha J_{0}^{(z)}}$ (see \ref{ap2} for computations) and afterwards by taking the limit $\alpha \rightarrow \infty$, we can map $H_{0}$ into a Hamiltonian $h_0$, such that its matrix representation is given in terms of triangular matrices:
\beqa
h_{0}=\frac{z}{2} \mu_{-} (J_{0}^{(z)})^2+ \sum_{n=1}^{N} a_{n} \left[ J_{0}^{(z)}\right]^{n}.
\label{h0}
\eeqa
As a consequence, we can state the following
\\

\noindent {\bf{Proposition 4.}}
The spectrum $\sigma(h_0)$ of $h_{0}$ is given by
\begin{equation}
\sigma(h_0)=\left\{
              \begin{array}{ll}
             \frac{z}{2} \mu_{-} (2k-1)^2 + u_{k}^{\pm} , & \hbox{ for } 1 \leq k \leq \frac{d}{2} ~ \mbox{and even }d\\
             \frac{z}{2} \mu_{-} 4(k-1)^2 +  v_{k}^{\pm}, & \hbox{ for } 1 \leq k \leq \frac{d+1}{2} ~\mbox{and odd }d
              \end{array}
            \right.
\label{spectrum2}    
\end{equation}
where
\begin{eqnarray*}
u_{k}^{\pm}&=&\sum_{n=1}^{N}(\pm 1)^{n} (2k-1)^{n}a_{n} \, ,\\
v_{k}^{\pm}&=&\sum_{n=1}^{N} (\pm 1)^{n} 2^{n+1} (k-1)^{n+1} a_{n} \, .
\end{eqnarray*}

Clearly, for $a_n \in \mathbb{R}$ the eigenvalues of $H_0$ of \eqref{hz1} belong to $\mathbb R$. It can be observed from \eqref{spectrum2} that for $a_n \ne 0$, 
the characteristic degeneracy of the spectrum of the operator $J_{-}$ is broken, giving rise again to bands of pairs of parallel lines separated by a controlled gap.  Note that the gap is symmetric when $u_{k}^{+}=u_{k}^{-}$ resp. $ (v_{k}^{+}=v_{k}^{-}) $, i.e when $a_{2n}=0$, for even (odd) dimension.
When odd coefficients are zero, i.e. all $a_{2n+1}=0$, we have $u_{k}^{+}=u_{k}^{-}$ (resp. $(v_{k}^{+}=v_{k}^{-})$) for even (odd) dimension, thus resulting in a Hamiltonian with degenerate spectrum.

\begin{figure}
\includegraphics[width=12cm]{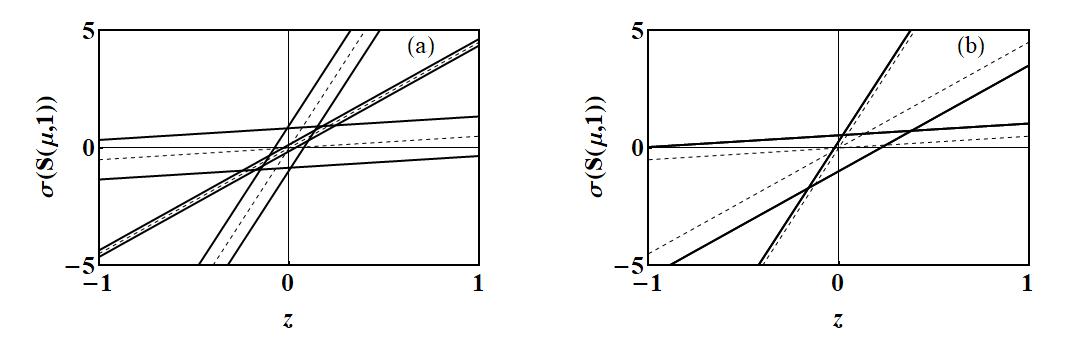}
\caption{Spectra of the Hamiltonians $S(\mu_{-}, \lambda)$ and $C(\mu_{-}, \lambda)$ in units of $\mu_{-}$, as a function of $z$, for dimension $d=6$ and $\lambda=1$ (solid lines). Dashed lines plot the spectrum of the Hamiltonian of \eqref{hz1}  for $a_n=0~\forall n$.}
\label{fig5}
\end{figure}

As an specific example, we can consider the Hamiltonians
\begin{equation}\label{sin}
S(\mu_{-}, \lambda)= \mu_{-} J_{-}^{(z)}+ \sin( \lambda J_{0}^{(z)}) \, ,
\end{equation}
\begin{equation}\label{cos}
C(\mu_{-}, \lambda)= \mu_{-} J_{-}^{(z)}+ \cos( \lambda J_{0}^{(z)}) \, .
\end{equation}
The operators $S(\mu_{-}, \lambda)$ and $C(\mu_{-}, \lambda)$ aare defined by power series with particular values of $a_n$, with even and odd null coefficients, respectively. 
In Figure \ref{fig5} we represent, with solid lines, the spectrum of Hamiltonians of \eqref{sin} and \eqref{cos}. We have plotted the case $\lambda=1$ for dimension $d=6$. As a guide, with dashed lines, we plot the spectrum of the Hamiltonian of \eqref{hz1} when $a_n=0~\forall n$. It can be seen from Panel (a) that, the spectrum of $S(\mu_{-}, 1)$  has pairs of parallel lines symmetrically separated with respect to the spectrum of $\mu_- (J_{-}^{(z)})^2$ by a controlled gap given by ${\pm\sin(1), \pm\sin(3),\pm\sin(5)}$, respectively. For $C(\mu_{-}, 1)$ we can see in Panel (b) that the degeneracy of $J_{-}^{(z)}$ is preserved, albeit displaced into a new double degenerate spectrum given by $\{\frac{z}{2}+\cos(1),\frac{9z}{2}+\cos(3),\frac{25z}{2}+\cos(5)\}$, due to the parity of the $\cos(x)$ function.

\section{Applications}\label{appli}

It is worth stressing that, recently, the separation in bands of parallel lines has been observed in the spectra of three electrons confined in an asymmetric two-dimensional double well, implemented by a two-centre-oscillator potential. This system turns out to be the cornerstone of two-dimensional (2D) semiconductor-based three-electron hybrid- double-quantum-dot (HDQD) qubits
(see \cite{ejQbit4,ejQbit3} and references therein). In the literature, theoretical model Hamiltonians have been developed to reproduce these experimental results \cite{ejQbits,ejQbits2}.  

The results presented in the previous Section strongly suggest the possibility of making use of specific $\pt$-symmetric Hamiltonians defined on the non-standard $U_z(sl(2,\mathbb R))$ quantum algebra in order to model these relevant systems. We show in the following that this will be indeed the case. 

Let us consider the effective Hamiltonian of \cite{ejQbit4} given by the equivalent form 
\beqa
H_{e}=\left(
\begin{array}{cccc}
 \delta L+\frac{\varepsilon }{2} & -t_{3} & 0 & t_{4} \\
 -t_{3} & -\frac{\varepsilon }{2}& t_1 & 0 \\
   0 & t_{1}& \frac{\varepsilon }{2} & -t_{2}\\
 t_{4} & 0 &  -t_{2} &\delta R-\frac{\varepsilon }{2} \\
\end{array}
\right),
\label{hqt}
\eeqa
where the parameter $\varepsilon$ models the detuning of three-electron hybrid qubits based on GaAs asymmetric double quantum dots, and with coupling constants $\delta L=3, \delta R= 95.8, t_{1}=1.8, t_2=7.1, t_3 =11.5, t_4 =6.3$ (in units of [GHz]) \cite{ejQbit3}. 
The eigenvalues of the Hamiltonian $H_{e}$ of \eqref{hqt} can be obtained analytically as the roots of a fourth-degree polynomial:
\begin{eqnarray}
p(\lambda)= \lambda^4+ c_3 \lambda^3+ c_2 \lambda^2 + c_1\lambda+ c_0.
\label{poly}
\end{eqnarray}
The explicit form of the coefficients $c_k$ is given in \ref{ap4}. For $\epsilon$ sufficiently large the eigenvalues of the Hamiltonian (\ref{hqt}) can be approximated by two sets of eigenvalues:
\beqa
E_{1,\pm} & = & \frac 12 \left ( \delta L \pm \sqrt{(\delta L+ \varepsilon)^2+ 4 t_3^2}\right), \nonn 
E_{2,\pm} & = & \frac 12 \left ( \delta R \pm \sqrt{(\delta R- \varepsilon)^2+ 4 t_2^2}\right).
\label{eigap}
\eeqa

An effective non-standard quantum algebra Hamiltonian $H_{eff}$ reproducing  the behaviour of the spectrum of $H_e$ (\ref{hqt}) an be obtained through
\beqa
H_{eff}= \left (
\begin{array}{cc} 
 1 & 0 \\
        0   & 0
\end{array}
\right)
\otimes H_{1} + 
\left ( \begin{array}{cc} 
 0 & 0 \\
        0   & 1
\end{array}
\right)
\otimes H_{2},
\label{heff}
\eeqa
with 
\beqa
H_1 & = & \frac 12 (\varepsilon+\delta L) J_0^{(\varepsilon)}+ \frac {t_3^2}{ \delta L} \varepsilon J_+^{(\varepsilon)}+ \frac{\delta L}{\varepsilon} J_-^{(\varepsilon)},\nonn
H_2 & = & \frac 12 (\varepsilon-\delta R) J_0^{(\varepsilon)}+ \frac {t_2^2}{ \delta R} \varepsilon J_+^{(\varepsilon)}+ \frac{\delta R}{\varepsilon} J_-^{(\varepsilon)}.
\label{effqdots}
\eeqa
by identifying the deformation parameter  with the detuning, therefore $z=\varepsilon$, and by making use of the two-dimensional irreducible representation of the $U_z(sl(2,\mathbb R))$ quantum algebra \eqref{2drep} obtained from \eqref{melements} with $\beta=-1$.

To obtain an isospectral Hamiltonian to $H_{eff}$, we construct the symmetry operator $S$ of \eqref{sym}, and from it the similarity transformation given by its  square root $S^{1/2}$: 
\beqa
h_{eff} & = & S^{1/2} \, H_{eff} \, S^{-1/2} \nonn
        & = & 
\left (
\begin{array}{cc} 
 h_1 & 0 \\
        0   & h_2 
\end{array}
\right),
\eeqa
being 
\beqa
S^{1/2} = 
\left (
\begin{array}{cc} 
 s_1 & 0 \\
        0   & s_2 
\end{array}
\right),
\eeqa
with 
\beqa
s_1 & = & 
\left (
\begin{array}{cc} 
 \frac{\varepsilon}{2 \delta L}\sqrt{-3 \delta L^2-2 \varepsilon \delta L + 4 t_3^2} & 0 \\
        0   & 1 
\end{array}
\right),
\nonumber \\
s_2 & = & 
\left (
\begin{array}{cc} 
 \frac{\varepsilon}{2 \delta R}\sqrt{\delta R^2-2 \varepsilon \delta R + 4 t_2^2} & 0 \\
        0   & 1 
\end{array}
\right) \, .
\eeqa
In this way we obtain that
\beqa
h_1 & = & 
\left (
\begin{array}{cc} 
-\frac12 (\delta L+ \varepsilon ) & \frac{1}{2}{ \bf i}\sqrt{-3 \delta L^2-2 \varepsilon \delta L + 4 t_3^2} \\
        -\frac{1}{2}{ \bf i}\sqrt{-3 \delta L^2-2 \varepsilon \delta L + 4 t_3^2}  & \frac12 (3 \delta L+\varepsilon )
\end{array}
\right),
\nonumber \\
h_2 & = & 
\left (
\begin{array}{cc} 
\frac12 (\delta R-\varepsilon ) & \frac{1}{2}{ \bf i}\sqrt{\delta R^2-2 \varepsilon \delta R + 4 t_2^2} \\
        -\frac{1}{2}{ \bf i}\sqrt{\delta R^2-2 \varepsilon \delta R + 4 t_2^2}   & \frac12 (\delta R+\varepsilon )
\end{array}
\right).
\eeqa
It is straightforward to prove that the eigenvalues of $h_1$ and $h_2$ are just $E_{1 \pm}$ and $E_{2 \pm}$, respectively.

Moreover, by making use of a second similarity transformation $P$, the Hamiltonian $h_{eff}$ can be arranged as
\beqa
\mathfrak{h} & =  & P \, h_{eff} \, P^{-1}, \nonn
& = & 
\left(
\begin{array}{cccc}
 \delta L+\frac{\varepsilon }{2} & -t_{3} & 0 & 0 \\
 -t_{3} & -\frac{\varepsilon }{2}& 0 & 0 \\
   0 & 0 & \frac{\varepsilon }{2} & -t_{2}\\
 0 & 0 &  -t_{2} &\delta R-\frac{\varepsilon }{2} \\
\end{array}
\right).
\eeqa
where
\beqa
P & = & 
\left (
\begin{array}{cc} 
 p_1 & 0 \\
        0   & p_2 
\end{array}
\right),
\eeqa
is given by
\beqa
p_1 & = & \left (
\begin{array}{cc} 
{\bf i} \frac{\varepsilon}{2 t_3} \sqrt{-3 \delta L^2-2 \varepsilon \delta L + 4 t_3^2} &  - \frac 1{2 t_3} (3 \delta L+2 \varepsilon )\\
0 & 1 
\end{array}
\right),\nonumber \\
p_2 & = & 
\left (
\begin{array}{cc} 
{\bf i} \frac{\varepsilon}{2 t_2} \sqrt{\delta R^2-2 \varepsilon \delta R + 4 t_2^2} &   \frac 1{2 t_2} (\delta R-2 \varepsilon )\\
0 & 1 
\end{array}
\right).
\eeqa

\begin{figure}
\includegraphics[width=13cm]{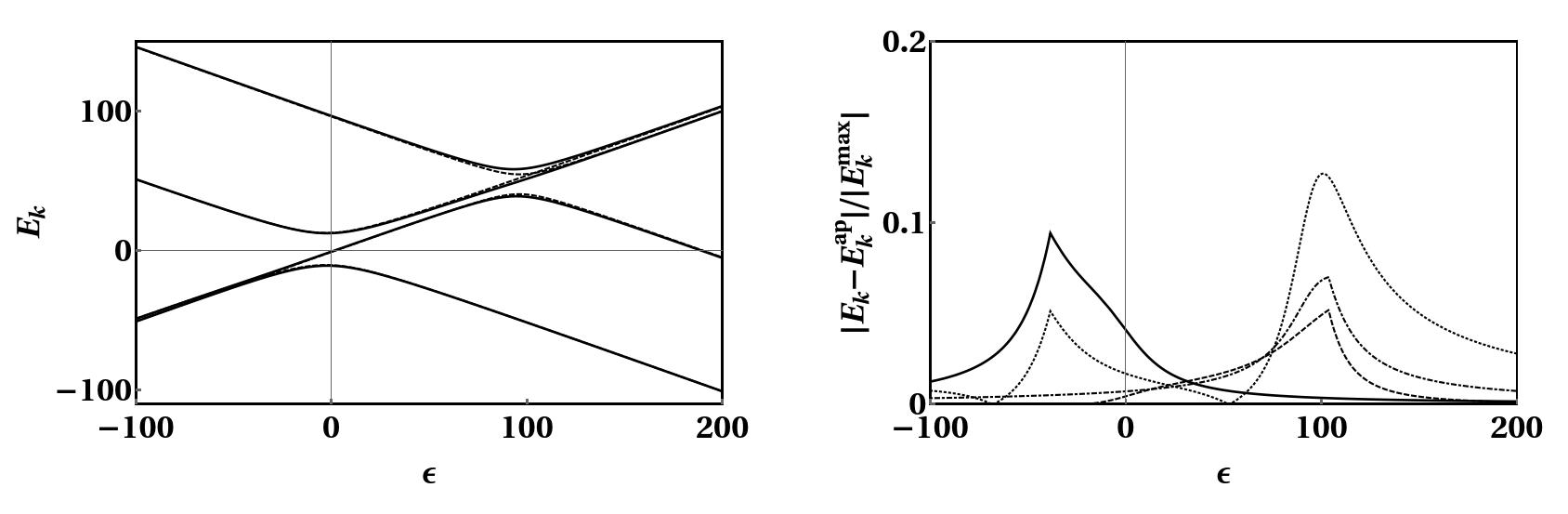}
\caption{The Figure depicts the spectrum of $H_e$ and $H_{eff}$ as a function of $\varepsilon$.In Panel (a), the exact eigenvalues of $H_e$ and the approximate values computed from \eqref{eigap} are displayed as a function of $\varepsilon$ (solid and dashed lines, respectively). In 
Panel (b), we plot the absolute value of the difference between the exact and the approximate eigenvalue in units of the maximum or minimum absolute value of the exact solution at the point where each band avoids the crossing. }
\label{fig8}
\end{figure}

Figure \ref{fig8} depicts the spectrum of $H_e$ and $H_{eff}$ as a function of $\varepsilon$.
In Panel (a), the exact eigenvalues of $H_e$ and their approximate values computed from 
(\ref{eigap}) are displayed as a function of $\varepsilon$ with solid and dashed lines, respectively. 
Panel (b) is devoted to analyse the differences between the energies deduced from the two models, which turn out to be very small under the identification between $\varepsilon$ and the deformation parameter $z$.

Therefore the previous example shows that realistic physical systems can be modeled by effective $\pt$-symmetry Hamiltonians constructed from the non-standard $U_z(sl(2,\mathbb R))$ algebra, and provided that the model parameters are chosen appropriately.

\section{Conclusions and outlook}\label{con}
In this work, we have obtained the analytical expression for the spectrum of a family of ${\mathcal {PT}}$-symmetric Hamiltonians defined in terms of the generators of the non-standard $U_{z}(sl(2, R))$ quantum algebra under a generic finite-dimensional irreducible representation of the latter \cite{uzfirst, ACC, BH1, BH2}. By generalising \cite{BH2}, we have presented a boson realisation of the generators 
of the $U_{z}(sl(2, R))$ algebra such that the co-product map and the commutation relations become invariant under the
$\mathcal{PT}$-transformation. In terms of these operators, 
we have introduced two families of ${\mathcal{PT}}$-symmetry Hamiltonians, given by (\ref{hz}) and (\ref{hz1}).

We have shown that the spectrum of the Hamiltonian $H$ in \eqref{hz} exhibits different properties depending on the relative signs of the parameters $\mu_\pm$. When ${\rm {sign}}(\mu_+)={\rm {sign}} ( \mu_-)$ the spectrum of $H$ of \eqref{hz} is real. Nevertheless, when ${\rm {sign}}(\mu_+)=-{\rm {sign}}(\mu_-)$, the spectrum of $H$ can include complex conjugate pairs of eigenvalues. Thus, we have two different dynamical phases, the exact ${\mathcal{PT}}$-symmetry phase for $\mu_0^2+ \mu_+ \mu_- >0$ with real energies, and the one for the broken $\pt$-symmetry phase for $\mu_0^2+ \mu_+ \mu_- <0$ consisting in pairs of complex conjugate eigenvalues. The boundary between these phases, given by $\mu_0^2+ \mu_+ \mu_- =0$, is formed by EPs. At these points, two or more eigenvalues are degenerated and their eigenvectors are coalescent.   

On the other hand, the spectrum of the Hamiltonian defined in \eqref{hz1} has been shown to consist, for real parameters, of real eigenvalues. As a characteristic feature of this spectrum, we have illustrated the appearance of bands consisting of pairs of eigenvalues, and we have studied the relation of the parameters of the model with the gap between such bands.

Remarkably enough, this particular band structure has suggested the definition of a non-standard quantum algebra effective model for the spectrum of a realistic system of three-electron hybrid qubits based on GaAs asymmetric double quantum dots \cite{ejQbit3}. In fact, by identifying the deformation parameter $z$ with the detuning $\varepsilon$ of the system, the spectrum of the effective Hamiltonian \eqref{effqdots} provides an excellent approximation to the energies of the actual physical system.

Work is in progress concerning the analytical spectra for more general ${\mathcal {PT}}$-symmetric Hamiltonians written in terms of the generators of the $U_{z}(sl(2, R))$ algebra. Also, their possible role as effective models for other quantum systems beyond the one here presented where the Hopf algebra deformation parameter $z$ had a neat physical interpretation.


\appendix

\section{}\label{ap2}

In what follows, we summarise the basic similarity transformations of the generators of  the $sl(2,\mathbb R)$ Lie algebra (\ref{sl2}):
\begin{eqnarray}
\re^{\alpha L_{+}}L_{-}\re^{-\alpha L_{+}}&=&  \alpha ( L_{0} - \alpha L_{+} )+ L_{-}, \nonn
\re^{\alpha L_{+}}L_{0}\re^{-\alpha L_{+}}&=& L_{0}- 2 \alpha L_{+}, \nonn
\re^{\alpha L_{-}}L_{+}\re^{-\alpha L_{-}}&=& - \alpha(L_{0} + \alpha L_{-})+ L_{+}, \nonn
\re^{\alpha L_{-}}L_{0}\re^{-\alpha L_{-}}&=& L_{0}+ 2 \alpha L_{-}, \nonn
\re^{\alpha L_{0}}L_{+}\re^{-\alpha L_{0}}&=& \re^{2 \alpha }L_{+} , \nonn
\re^{\alpha L_{0}}L_{-}\re^{-\alpha L_{0}}&=& \re^{-2 \alpha }L_{-} \, .
\end{eqnarray}
For the $U_z(sl(2,\mathbb R))$ quantum algebra (\ref{genpt}) we have
\begin{eqnarray}
\re^{\alpha J_{+}^{(z)}}   J_{-}^{(z)}   \re^{-\alpha J_{+}^{(z)}} &=&  \alpha \left( J_{0}^{(z)} - \alpha f(J_{+}^{(z)}) \right)+ J_{-}^{(z)}, \nonn
 \re^{\alpha J_{+}^{(z)}}  J_{0}^{(z)}   \re^{-\alpha J_{+}^{(z)}} &=& J_{0}^{(z)}- 2 \alpha f(J_{+}^{(z)}), \nonn
 \re^{\alpha J_{-}^{(z)}}  J_{0}^{(z)}   \re^{-\alpha J_{-}^{(z)}} &=& -d_{\alpha } \left( \re^{\alpha J_{-}^{(z)}} J_{+}^{(z)} \re^{-\alpha J_{-}^{(z)}} \right), \nonn
\re^{\alpha J_{0}^{(z)}}  J_{+}^{(z)}   \re^{-\alpha J_{0}^{(z)}} &=& 
 \sum_{n=1}^{\infty} \frac{(-2z)^{n-1}}{n} \left (1-(-2 \re^{\alpha}\sinh(\alpha))^{n})(f(J_{+}^{(z)}) \right)^{n}, \nonn
\re^{\alpha J_{0}^{(z)}}   J_{-}^{(z)}  \re^{-\alpha J_{0}^{(z)}}  &=& \re^{-2 \alpha }J_{-}^{(z)}+ z \re^{-\alpha} \sinh(\alpha) J_{z}^{(z)2}, \nonn
\re^{\alpha J_{0}^{(z)}}   f(J_{+}^{(z)})  \re^{-\alpha J_{0}^{(z)}}&=& 
\sum_{n=1}^{\infty} \re^{\alpha} \left(4 z \sinh(\alpha) \right)^{n-1} \left(\re^{\alpha} f(J_{+}^{(z)})\right)^{n},
\end{eqnarray}
where the function $f$ is given by
\beqa
f(J_{+}^{(z)})= \frac 12 ~ J_{+}^{(z)}~[J_{0}^{(z)},J_{+}^{(z)}].
\eeqa

\section{}\label{ap1}

We shall prove that the coproduct ($\Delta$), the counit ($\varepsilon$), antipode ($\gamma$ ) maps and the commutation rules amongst the operators $\{{J}_{0}^{(z)},{J}_{+}^{(z)},{J}_{-}^{(z)} \}$ have the same structure as those of 
$\{{j}_{0}^{(z)},{j}_{+}^{(z)},{j}_{-}^{(z)} \}$. 

Let us start with the Hopf structure of the operators $\{{j}_{0}^{(z)},{j}_{+}^{(z)},{j}_{-}^{(z)} \}$:
\begin{eqnarray}
\Delta ({    {j}}_0^{(z)}) & = & 1 \otimes {    {j}}_0^{(z)} + {    {j}}_0^{(z)} \otimes \re^{2 z {    {j}}_+^{(z)} }, \nonumber \\
\Delta ({    {j}}_-^{(z)}) & = & 1 \otimes {    {j}}_-^{(z)} + {    {j}}_-^{(z)} \otimes \re^{2 z {    {j}}_+^{(z)} }, \nonumber \\
\Delta ({    {j}}_+^{(z)}) & = & 1 \otimes {    {j}}_+^{(z)} + {    {j}}_+^{(z)} \otimes 1, \nonumber \\
\varepsilon(X)& = & 0, ~~~X \in \{     {j}_{0}^{(z)},    {j}_{+}^{(z)},     {j}_{-}^{(z)} \}, \nonumber \\
\gamma (    {j}_{0}^{(z)}) &=& -{    {j}}_{0}^{(z)} \re^{-2 z {    {j}}_{+}^{(z)}}, \nonumber \\
\gamma (    {j}_{-}^{(z)}) &=& -{    {j}}_{-}^{(z)} \re^{-2 z     {j}_{+}^{(z)} }, \nonumber \\
\gamma (    {j}_{+}^{(z)}) &=& -{    {j}}_{+}^{(z)},
\end{eqnarray}
we shall write the maps for $\{{J}_{0}^{(z)},{J}_{+}^{(z)},{J}_{-}^{(z)} \}$ in terms of $\{{j}_{0}^{(z)},{j}_{+}^{(z)},{j}_{-}^{(z)} \}$:
\begin{eqnarray}
\Delta( {J}_{0}^{(z)})&=& \Delta( {j}_{0}^{(-\uni z)})=
 1 \otimes {j}_{0}^{(-\uni z)} +{j}_{0}^{(-\uni z)} \otimes \re^{2 (- \uni z) (j_{+}^{(- \uni z)})} \nonumber \\ &= &
 1 \otimes {J}_{0}^{(z)} +{J}_{0}^{(z)} \otimes \re^{2 {z} {J}_{+}^{(z)}}, \nonumber \\
\Delta ({J}_{-}^{(z)}) & = & \Delta (\uni {j}_{-}^{( -\uni z)})= 1 \otimes \uni j_{-}^{(-\uni z)} + \uni j_{-}^{(-\uni z)} \otimes  \re^{2 (- \uni z) (j_{+}^{(- \uni z)})} \nonumber \\
&=&1 \otimes{J}_{-}^{(z)} +{J}_{-}^{(z)} \otimes \re^{2 {z} {J}_{+}^{(z)}}, \nonumber \\
\Delta ( {J}_{+}^{(z)}) & = & 1 \otimes (- \uni j_{+}^{(-\uni z)})+(- \uni j_{+}^{(-\uni z)}) \otimes 1  \nonumber \\
&=&1 \otimes {J}_{+}^{(z)} +{J}_{+}^{(z)} \otimes 1, \nonumber \\
\varepsilon(X)& = & 0, ~~~X \in \{     {J}_{0}^{(z)},    {J}_{+}^{(z)},     {J}_{-}^{(z)} \}, \nonumber \\
\gamma ( {J}_{0}^{(z)}) &=& \gamma ( {j}_{0}^{(- \uni z)})= -{{j}}_{0}^{(-\uni z)} \re^{-2 (-\uni z)  {j}_{+}^{(-\uni z)}}
\nonumber \\
&= & -J_0^{(z)} \re^{-2 z J_+^{(z)}}, \nonumber \\
\gamma ( {J}_{-}^{(z)}) &=& \gamma (\uni {j}_{-}^{(-\uni z)})=-{\uni{j}}_{-}^{(-\uni z)} \re^{-2 (-\uni z)  {j}_{+}^{(-\uni z)}} \nonumber \\
& = & -{J}_{-}^{(z)} \re^{-2 z J_+^{(z)}}, \nonumber \\ 
\gamma ({J}_{+}^{(z)}) &=& \gamma (-\uni {j}_{+}^{(-\uni z)})= -{ (-\uni )   {j}}_{+}^{(-\uni z)} \nonumber \\
& = & -{J}_{+}^{(z)}\, .
\end{eqnarray}
For the commutation relations, we have:
\begin{eqnarray}\
~ [J_{0}^{(z)},J_{+}^{(z)}]&=&[j_{0}^{(- \uni z)},-\uni j_{+}^{(-\uni z)}]\nonumber  \\
&=&-\uni \tfrac{e^{2 (- \uni z) j_{+}^{(-\uni z)}}-1}{-\uni z} \nonumber  \\ 
&=&\tfrac{e^{2 z J_{+}^{(z)}}-1}{z}, \nonumber  \\
~ [ J_{0}^{(z)},J_{-}^{(z)} ] &=& [ j_{0}^{(- \uni z)},\uni j_{-}^{(\uni z)}]\nonumber  \\
&=&\uni(- 2 j_{-}^{(-\uni z)}+(-\uni z) (j_{0}^{(-\uni z)})^{2} )\nonumber  \\
&=&- 2 \uni j_{-}^{(-\uni z)}+z (j_{0}^{(-\uni z)})^{2} \nonumber  \\
&=&- 2 J_{-}^{(z)}+z (J_{0}^{(z)})^{2} , \nonumber \\
~ [J_{+}^{(z)},J_{-}^{(z)}]&=&[-\uni j_{+}^{(-\uni z)},-\uni j_{-}^{(-\uni z)}] \nonumber \\
&=& j_{0}^{(-\uni z)} \nonumber \\
&=&J_{0}^{(z)}.
\end{eqnarray}

Next, we shall prove the invariance of the coproduct and the commutation relations under a $\pt$ symmetry transformation. Let us summarise the transformation properties of the different operators and scalars under $\pt$-symmetry:
\beqa
j_0^{(z)} & \rightarrow & ~~~j_0^{(-z)}, \nonumber \\
j_\pm^{(z)} & \rightarrow & -j_\pm^{(-z)}, \nonumber \\
\uni & \rightarrow & -\uni, \nonumber \\
    {J}_0^{(z)} & \rightarrow &     {J}_0^{(z)}, \nonumber \\
    {J}_\pm^{(z)} & \rightarrow &     {J}_\pm^{(z)}.
\eeqa
Therefore we have: 
\begin{eqnarray}
({\mathcal{PT}})\Delta (    {J}_{0}^{(z)}) (\pt)^{-1}&=& (\pt) ( 1 \otimes {    {J}}_0^{(z)} + {    {J}}_0^{(z)} \otimes \re^{2 z {    {J}}_+^{(z)} }) (\pt)^{-1} 
\nonumber \\ 
& = & 1 \otimes (    {J}_{0}^{(z)}) +({    {J}_{0}^{(z)}}) \otimes \re^{ 2 z(    {J}_{+}^{(z)})}
\nonumber \\
& = & \Delta (    {J}_{0}^{(z)}), \nonumber \\
({\mathcal{PT}})\Delta (    {J}_{-}^{(z)}) (\pt)^{-1}&=& (\pt) ( 1 \otimes {    {J}}_-^{(z)} + {    {J}}_-^{(z)} \otimes \re^{2 z {    {J}}_+^{(z)} }) (\pt)^{-1} 
\nonumber \\ 
& = & 1 \otimes (    {J}_{-}^{(z)}) +({    {J}_{-}^{(z)}}) \otimes \re^{2 z(    {J}_{+}^{(z)})}
\nonumber \\
& = & \Delta (    {J}_{-}^{(z)}), \nonumber \\
({\mathcal{PT}})\Delta (    {J}_{+}^{(z)}) (\pt)^{-1}&=& ( \pt) ( 1 \otimes {    {J}}_+^{(z)} + {    {J}}_+^{(z)} \otimes 1 ) (\pt)^{-1} 
\nonumber \\ 
& = & 1 \otimes (    {J}_{+}^{(z)}) +({    {J}_{+}^{(z)}}) \otimes 1
\nonumber \\
& = & \Delta (    {J}_{+}^{(z)}), \nonumber \\
\end{eqnarray}

In a similar way, we can show that the commutation relations are also invariant under $\pt$-symmetry transformations:
\beqa
(\pt) [     {J}_{0}^{(z)},    {J}_{+}^{(z)}] (\pt)^{-1} & = & (\pt) \left( \frac{\re^{2 z     {J}_{+}^{(z)}}-1}{z} \right) (\pt)^{-1} \nonumber \\
& = &  \frac{\re^{2 z    {J}_{+}^{(z)}}-1}{z}\nonumber \\
&= & [     {J}_{0}^{(z)},    {J}_{+}^{(z)}]
\nonumber \\
(\pt) [    {J}_{0}^{(z)},    {J}_{-}^{(z)}]  (\pt)^{-1} & = & (\pt) (-2     {J}_{-}^{(z)}+z \left(     {J}_{0}^{(z)} \right)^2) (\pt)^{-1} \nonumber \\
& = & -2     {J}_{-}^{(z)}+z \left(     {J}_{0}^{(z)} \right)^2
\nonumber \\
&= & [    {J}_{0}^{(z)},    {J}_{-}^{(z)}],\nonumber \\
(\pt) [    {J}_{+}^{(z)},    {J}_{-}^{(z)}]  (\pt)^{-1} & = & (\pt) (    {J}_{0}^{(z)}) (\pt)^{-1}, \nonumber \\
& = &     {J}_{0}^{(z)}\nonumber \\
&= & [    {J}_{+}^{(z)},    {J}_{-}^{(z)}].
\eeqa

\section{}\label{ap4}
 The coefficients of the characteristic polynomial $p(\lambda)$ of Eq.(\ref{poly}) are given by: 
\begin{eqnarray}
c_3(\varepsilon) &=& -\delta_{L}-\delta_{R}, \nonn
c_2(\varepsilon) &=& \frac{1}{2} \left(\varepsilon 
   (\delta_{R}-\delta_{L})-2 \left(-\delta_{L} \delta_{R}+t_{1}^2+t_{2}^2+t_{3}^2+t_{4}^2\right)-\varepsilon ^2\right), \nonn
c_1(\varepsilon) &=& \frac{1}{4}
   \varepsilon ^2 (\delta_{L}+\delta_{R})+t_{1}^2 (\delta_{L}+\delta_{R})+\delta_{L} t_{2}^2+\delta_{R} t_{3}^2, \nonn
c_0(\varepsilon) &=& \frac{1}{16} 
\left( \varepsilon^4 +
2 \varepsilon ^3 (\delta_{L}-\delta_{R})
+ 4 \varepsilon ^2 \left(\delta_{L}\delta_{R}+t_{1}^2+t_{2}^2+t_{3}^2+t_{4}^2\right)
\right. \nonn
& & \left. 
+8 \varepsilon \left(\delta_{L} \left(t_{1}^2+t_{2}^2\right) -\delta_{R} \left(t_{1}^2+t_{3}^2\right) \right)
+16 \left((t_{2}t_{3}-t_{1}
   t_{4})^2-\delta_{L} \delta_{R} t_{1}^2\right) \right).\nonumber  
\end{eqnarray}  
The exact expression for the roots of the quartic equation $p(\lambda)=0$ can be found in \cite{abra}.

\section*{Acknowledgements}

A.B. has been partially supported by Agencia Estatal de Investigaci\'on (Spain)  under grant  PID2019-106802GB-I00/AEI/10.13039/501100011033, and by the Q-CAYLE Project funded by the Regional Government of Castilla y Le\'on (Junta de Castilla y Le\'on) and by the Ministry of Science and Innovation MICIN through the European Union funds NextGenerationEU (PRTR C17.I1). M.R. is grateful to
the Universidad de Burgos for its hospitality. M.R. and R.R. have been partially supported by the grant 11/X982 of the University of La Plata (Argentine).


\end{document}